\documentclass[aip,reprint]{revtex4-1}
\usepackage{amsmath}
\usepackage{amssymb}
\usepackage{amsthm}
\usepackage{fancyhdr}
\usepackage{graphicx}
\usepackage{hyperref}
\usepackage{multirow}
\usepackage{rotating}
\usepackage{float}
\usepackage{array}
\usepackage{supertabular}
\usepackage{amsmath}
\usepackage{dcolumn}
\usepackage{tipa}
\usepackage{csquotes}
\usepackage{siunitx}
\DeclareSIUnit{\calorie}{cal}
\DeclareSIUnit\atm{atm}
\usepackage{tikz}
\usetikzlibrary{decorations.pathmorphing}
\usetikzlibrary{shapes.geometric}
\usetikzlibrary{arrows.meta, positioning}
\tikzset{
bcm/.style = {circle, fill=yellow, minimum size=#1,
              inner sep=0pt, outer sep=0pt},
bcm/.default = 6.5pt
}
\tikzset{
  half circle/.style={
      semicircle,
      shape border rotate=90,
      anchor=chord center,
      minimum size=3mm
      }
}
\tikzset{
  half circle alt/.style={
      semicircle,
      shape border rotate=90,
      anchor=chord center,
      minimum size=6.2mm
      },
   arrow/.style = {
    draw=blue,
    line width=0.5mm,
    -{Triangle[length=3mm,width=1.5mm]},
    shorten >=1mm, shorten <=1mm,
    font=\fontsize{8}{8}\selectfont}
}
\tikzset{cross/.style={cross out, draw=red, minimum size=5*(#1-\pgflinewidth), inner sep=0pt, outer sep=1pt},
cross/.default={2pt}}
\usetikzlibrary{shapes,backgrounds}
\usepackage{tkz-euclide}

\usepackage{pgfplots}
\pgfplotsset{ compat=1.18,}
\usepgfplotslibrary{colormaps, groupplots}
\usetikzlibrary{decorations.pathmorphing}
\tikzset{
bcm/.style = {circle, fill=yellow, minimum size=#1,
              inner sep=0pt, outer sep=0pt},
bcm/.default = 6pt 
}
\tikzset{
    set/.style={
        execute at end picture={
            \pgfextractx{\dimen0}{\pgfpointanchor{current bounding box}{south west}}
            \xdef\savedxmin{\the\dimen0}
            \pgfextractx{\dimen0}{\pgfpointanchor{current bounding box}{north east}}
            \xdef\savedxmax{\the\dimen0}
        }
    },
    receive/.style={
        execute at end picture={
            \pgfextracty{\dimen0}{\pgfpointanchor{current bounding box}{south west}}
            \edef\localymin{\the\dimen0}
            \pgfextracty{\dimen0}{\pgfpointanchor{current bounding box}{north east}}
            \edef\localymax{\the\dimen0}
            \pgfresetboundingbox
            \path (\savedxmin,\localymin) rectangle (\savedxmax,\localymax);
        }
    }
}
\usetikzlibrary{calc}

\DeclareSIUnit\torr{Torr}
\usepackage{textgreek}

\begin{document}

\title{Fabry-P\'erot Resonance Shifts as a Novel Non-destructive In Situ Surface Damage Diagnostic}

\author{G. Blume}
\email{gtblume@jlab.org}
\affiliation{Department of Physics, Old Dominion University, Norfolk, Virginia 23529, USA}

\author{C. Kirk}
\affiliation{Department of Electrical and Computer Engineering, Old Dominion University, \\Norfolk, Virginia 23529, USA}

\author{S. Poudel}
\affiliation{Department of Electrical and Computer Engineering, Old Dominion University, \\Norfolk, Virginia 23529, USA}

\author{J. Hill}
\affiliation{Department of Electrical and Computer Engineering, Old Dominion University, \\Norfolk, Virginia 23529, USA}

\author{A. Kachwala}
\affiliation{Thomas Jefferson National Accelerator Facility, Newport News, VA 23606, USA}

\author{J. Grames}
\affiliation{Thomas Jefferson National Accelerator Facility, Newport News, VA 23606, USA}

\author{M. L. Stutzman}
\affiliation{Thomas Jefferson National Accelerator Facility, Newport News, VA 23606, USA}

\author{M. Grau}
\affiliation{Department of Physics, Old Dominion University, Norfolk, Virginia 23529, USA}

\author{S. Marsillac}
\affiliation{Department of Electrical and Computer Engineering, Old Dominion University, \\Norfolk, Virginia 23529, USA}

\date{\today}

\begin{abstract}
Current surface damage evaluation for nanostructured electronic devices requires specialized equipment or conditions outside the designed operating environment and cannot be used for in situ evaluation of surface damage.
In this work, we developed a non-destructive method to probe the surface sublimation of a semiconductor heterostructure which can be applied to other semiconductor damage mechanisms.
This approach employs shifts in the resonances of an optical cavity comprising the surface and an integrated distributed Bragg reflector fabricated below the semiconductor heterostructure undergoing sublimation.
We performed the demonstration of this technique under vacuum, but surface damage effects can be observed in any operational environment either in situ or ex situ.
\end{abstract}

\maketitle

Nanostructured strained superlattice (SSL) GaAs based photocathodes are used for the electron beam production in spin-polarized accelerator programs~\citep{Adderly12gev,joesgaas}.
Two fabrication techniques exist for these photocathodes: molecular beam epitaxy (MBE)~\citep{wei,mbebnl,maruyama1,maryyama2} and metal-organic chemical vapor deposition (MOCVD)~\citep{belfore,masters}.
For both techniques, understanding the surface condition is paramount to successful design and operation~\citep{biswas}.
Current damage evaluation of these devices requires removal from the operating environment and are destructive~\citep{TEMstuff,FIBstuff,ATMstuff} or require expensive equipment with specific vacuum conditions~\citep{ellipsometry,stm} that do not mimic regular operation.
However, a solution exists using the in situ method to measure Fabry-P\'erot (FP) resonance shifts to observe thickness changes~\citep{tech}.

In this work, we used the wavelength dependent reflectivity to measure surface changes of both a MOCVD-grown and MBE-grown SSL GaAs photocathodes during heat exposure at \SI{550}{\degreeCelsius}. 
At this temperature, GaAs will sublimate along with any native oxides and implanted material, changing the surface quality and sample thickness~\citep{zhou2010,GOLDSTEIN1976733}.
For both photocathodes, the SSL was deposited atop a distributed Bragg reflector (DBR) which consists of a superlattice of materials with alternating refractive indices that reflect light for enhanced photon absorption over a narrow range of wavelengths~\citep{Saka_1993}.
The surface and DBR form an optical cavity whose resonances are wavelength dependent and determined, in part, by the distance between the DBR and the surface~\citep{ZHMAKIN2011189}.
Sublimation of the surface layers reduces this distance and shifts the resonance to shorter wavelengths at a rate proportional to the rate of sublimation.
This resonance shift rate, as a function of heat treatment, was used to extract the rate of sublimation using a transfer matrix method (TMM)~\citep{Byrnes2016MultilayerOC} calculations to convert resonance shifts to changes in material thickness.
Scanning transmission electron microscopy (STEM) imaging confirmed this rate.
This approach measured an average sublimation rate of \SI{0.94(0.03)}{\nano\meter\per\hour} for the MOCVD sample and \SI{0.93(0.01)}{\nano\meter\per\hour} for the MBE sample at \SI{550}{\degreeCelsius}.
Finally, we discuss the implications on photocathode performance due to the shift in resonant wavelength. 

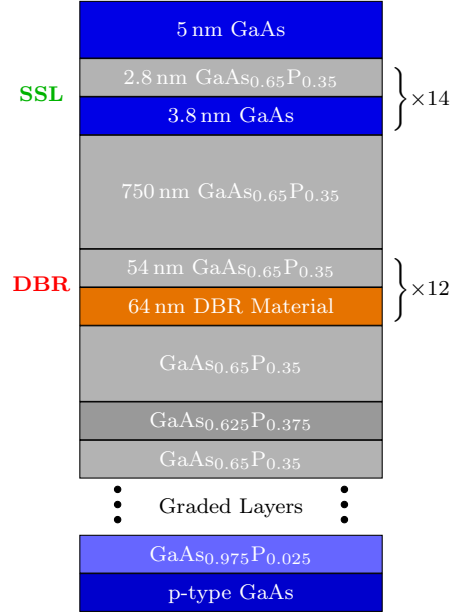
\begin{figure}[htb!]\centering%
\begin{tikzpicture}
[
    >=latex,
    photon/.style={red, decorate, decoration={snake, post length=1mm}, ->},
    annotation/.style={font=\footnotesize},
]
\draw (-5.5, 0) node[draw, fill=blue!80!black, rectangle, rounded corners=0mm, minimum width=4cm, minimum height=0.5cm, inner sep=0] (lbase) {};
\draw (lbase.south) ++ (0,0.48) node[anchor=north, annotation, color=white] {p-type GaAs};

\draw (lbase.north)++(0,0.25) node[draw, fill=blue!60!white, rectangle, rounded corners=0mm, minimum width=4cm, minimum height=0.5cm, inner sep=0] (l2) {};
\draw (l2.south) ++ (0,0.48) node[anchor=north, annotation, color=white] {$\text{GaAs}_{0.975}\text{P}_{0.025}$};

\draw (l2.north)++(0,1) node[draw, fill=white!70!black, rectangle, rounded corners=0mm, minimum width=4cm, minimum height=0.5cm, inner sep=0] (l3) {};
\draw (l3.south) ++ (0,0.48) node[anchor=north, annotation, color=white] {$\text{GaAs}_{0.65}\text{P}_{0.35}$};

\draw (l3.north)++(0,0.25) node[draw, fill=white!60!black, rectangle, rounded corners=0mm, minimum width=4cm, minimum height=0.5cm, inner sep=0] (l4) {};
\draw (l4.south) ++ (0,0.48) node[anchor=north, annotation, color=white] {$\text{GaAs}_{0.625}\text{P}_{0.375}$};

\draw (l4.north)++(0,0.5) node[draw, fill=white!70!black, rectangle, rounded corners=0mm, minimum width=4cm, minimum height=1cm, inner sep=0] (l5) {};
\draw (l5.south) ++ (0,0.73) node[anchor=north, annotation, color=white] {$\text{GaAs}_{0.65}\text{P}_{0.35}$};

\draw (l5.north)++(0,0.25) node[draw, fill=orange!90!black, rectangle, rounded corners=0mm, minimum width=4cm, minimum height=0.5cm, inner sep=0] (l6) {};
\draw (l6.south) ++ (0,0.48) node[anchor=north, annotation, color=white] {\SI{64}{\nano\meter} DBR Material};

\draw (l6.north)++(0,0.25) node[draw, fill=white!70!black, rectangle, rounded corners=0mm, minimum width=4cm, minimum height=0.5cm, inner sep=0] (l7) {};
\draw (l7.south) ++ (0,0.48) node[anchor=north, annotation, color=white] {\SI{54}{\nano\meter} $\text{GaAs}_{0.65}\text{P}_{0.35}$};

\draw (l7.north)++(0,0.75) node[draw, fill=white!70!black, rectangle, rounded corners=0mm, minimum width=4cm, minimum height=1.5cm, inner sep=0] (l8) {};
\draw (l8.south) ++ (0,0.98) node[anchor=north, annotation, color=white] {\SI{750}{\nano\meter} $\text{GaAs}_{0.65}\text{P}_{0.35}$};

\draw (l8.north)++(0,0.25) node[draw, fill=blue!90!black, rectangle, rounded corners=0mm, minimum width=4cm, minimum height=0.5cm, inner sep=0] (l9) {};
\draw (l9.south) ++ (0,0.48) node[anchor=north, annotation, color=white] {\SI{3.8}{\nano\meter} GaAs};

\draw (l9.north)++(0,0.25) node[draw, fill=white!70!black, rectangle, rounded corners=0mm, minimum width=4cm, minimum height=0.5cm, inner sep=0] (l10) {};
\draw (l10.south) ++ (0,0.48) node[anchor=north, annotation, color=white] {\SI{2.8}{\nano\meter} $\text{GaAs}_{0.65}\text{P}_{0.35}$};

\draw (l10.north)++(0,0.375) node[draw, fill=blue!90!black, rectangle, rounded corners=0mm, minimum width=4cm, minimum height=0.75cm, inner sep=0] (l11) {};
\draw (l11.south) ++ (0,0.61) node[anchor=north, annotation, color=white] {\SI{5}{\nano\meter} GaAs};

\draw (l7.east) ++ (0.5,0.25) node[anchor=north, annotation, color=black] {\Bigg\}$\times$12};
\draw (l10.east) ++ (0.5,0.25) node[anchor=north, annotation, color=black] {\Bigg\}$\times$14};
\draw (l7.west) ++ (-0.5,0.0) node[anchor=north, annotation, color=red] {\textbf{DBR}};
\draw (l10.west) ++ (-0.5,0.0) node[anchor=north, annotation, color=green!70!black] {\textbf{SSL}};

\draw (l2.north) ++ (0,0.605) node[anchor=north, annotation, color=black] {Graded Layers};

\draw[black,fill=black] (l2.north)++(1.5,0.6) circle[radius=1.2pt];
\draw[black,fill=black] (l2.north)++(1.5,0.4) circle[radius=1.2pt];
\draw[black,fill=black] (l2.north)++(1.5,0.2) circle[radius=1.2pt];
\draw[black,fill=black] (l2.north)++(-1.5,0.6) circle[radius=1.2pt];
\draw[black,fill=black] (l2.north)++(-1.5,0.4) circle[radius=1.2pt];
\draw[black,fill=black] (l2.north)++(-1.5,0.2) circle[radius=1.2pt];

\end{tikzpicture}
\caption{Diagram of the samples used in this sublimation study. Both the MOCVD and MBE sample contain a 12 pair DBR and 14 pair SSL. The DBR material for the MOCVD sample was $\text{In}_{0.30}\text{Al}_{0.70}$P and for the MBE sample was $\text{Al}\text{As}_{0.60}\text{P}_{0.40}$. The optical cavity consists of material located between the surface and the top of the DBR.}
\label{fig:prin2}
\end{figure}

The structure of both the MOCVD and MBE samples evaluated in this work begin with a p-type doped GaAs substrate upon which a metamorphic buffer layer is deposited. 
This begins with an initial phosphorus composition of \SI{0}{\percent} and ends with a composition of \SI{35}{\percent}.  
All these layers are p-doped with Zn at $5 \times 10^{18}$ \SI{}{\centi\meter}$^{-3}$.

Following the metamorphic layer is a buffer layer and then the DBR. 
The DBR consists of 12 pairs of staggered material with flipped high and low refractive indices to create an optical mirror, enabling reflected light to be absorbed in the SSL located above. 
In addition to the $\text{GaAs}_{0.65}\text{P}_{0.35}$ the DBR material was $\text{In}_{0.30}\text{Al}_{0.70}$P for MOCVD and $\text{Al}\text{As}_{0.60}\text{P}_{0.40}$ for MBE.
The DBR is p-doped at $5 \times 10^{18}$ \SI{}{\centi\meter}$^{-3}$.

Next there is a \SI{750}{\nano\meter} thick spacer before 14 pairs of a photoemmisive strained superlattice consisting of in-plane compressive strained GaAs and un-strained $\text{GaAs}_{0.65}\text{P}_{0.35}$. 
The nominal thicknesses are \SI{3.8}{\nano\meter} and \SI{2.8}{\nano\meter} respectively each doped p-type at $5 \times 10^{17}$ \SI{}{\centi\meter}$^{-3}$. 
Finally, a highly carbon doped ($5 \times 10^{19}$ \SI{}{\centi\meter}$^{-3}$) strained GaAs layer is deposited on the surface. 
Figure \ref{fig:prin2} depicts the structure for both samples. 

The congruent evaporation temperature is the highest temperature in which solid GaAs sublimates and the vapor products have the same chemical composition as the solid.  
Above the congruent evaporation temperature of roughly \SI{620}{\degreeCelsius}~\citep{zhou2010} for GaAs, arsenic is preferentially sublimated from the surface leaving behind gallium droplets in a process frequently referred to as frosting~\citep{lowes,allmang,clayburn}. 
However, below this temperature, sublimation continues while maintaining the stoichiometric ratio between Ga and As of 1:1~\citep{zhou2010,GOLDSTEIN1976733}.
In this regime of congruent sublimation, the sublimation rate is determined by the equilibrium pressure of gallium vapor~\citep{THURMOND1965785,ARTHUR19672257}.
With no gallium overpressure, GaAs will sublimate and produce a chamber pressure equal to the equilibrium pressure which we used to calculate the sublimation rate in the Langmuir model~\citep{Langmuir1913VaporPressure}.
Removal of surface material changes the thickness of the optical cavity and therefore the resonance of the DBR causing the wavelength of decreased reflectivity to shift.
We measured this shift and used it to calculate the material removed in situ.

Langmuir sublimation describes the vaporization of a solid material into vacuum~\citep{Langmuir1913VaporPressure}.
Here, the rate of mass loss is described as
\begin{equation}
    Z = \alpha (P_{\text{GaAs}}(T)-P_0)\sqrt{\frac{M}{2\pi RT}},
\end{equation}
where $\alpha=1$ is the Langmuir coefficient, $P_{\text{GaAs}}(T)$ is the equilibrium vapor pressure of gallium, which is equal to the vapor pressure of gallium below the congruent temperature, $P_0$ is the chamber background pressure, $M$ is the molar mass of GaAs, $R=$\SI{1.9872}{\calorie\per\mole\per\kelvin} is the universal gas constant, and $T$ is temperature.
The Langmuir coefficient describes the percentage of material that will sublimate and not recombine with the surface.
Langmuir sublimation is a useful model in the temperature range where congruent sublimation occurs as the material composition remains unchanged.

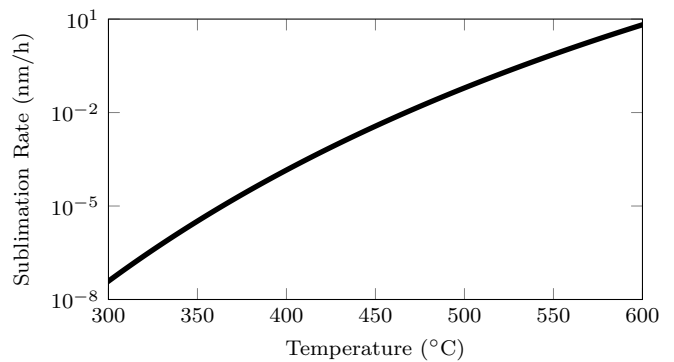
\begin{figure}[ht!]
    \begin{tikzpicture}[set]
\begin{axis}
[
    legend style={font=\small},
    legend pos=north west, 
    width=1\linewidth,
    height=.612\linewidth,
    xlabel={Temperature (\SI{}{\degreeCelsius})},
    ylabel={Sublimation Rate (\SI{}{\nano\meter\per\hour})},
    ymode=log,
    ymin=1e-8, ymax=10,
    xmin=300, xmax=600,
    tick label style={font=\footnotesize},
    label style={font=\footnotesize},
    name=border
]
\addplot+[no marks, black, line width=2pt, mark options={fill=black, draw=black}] table[x, y index=1] {data/expectedsubrate.dat};

\end{axis}
\end{tikzpicture}
\caption{The temperature dependent expected sublimation rate of GaAs from Langmuir sublimation.}
\label{expectedsubrate}
\end{figure}

The pressure produced in this regime corresponds to the equilibrium pressure of gallium~\citep{ARTHUR19672257} which we calculated as a function of temperature in \SI{}{\kelvin} via
\begin{equation}
    \text{log}(P) = A + \frac{B}{T} + C\cdot\text{log}(T),
    \label{pressureeq}
\end{equation}
for pressure in atmospheres, where the constants $A$=$6.754$, $B$=$-13984$, and $C$=$-0.3413$ are tabulated for gallium by C. B. Alcock \textit{et al.} ~\citep{Alcock01071984}.
At \SI{550}{\degreeCelsius}, Eq. \ref{pressureeq} gives a gallium equilibrium pressure of ($6.0 \pm 0.3 \times10^{-7} $) \SI{}{\pascal} leading to an expected sublimation rate, $R_d$, of over \SI{0.5}{\nano\meter\per\hour}.
The rate of mass loss can be converted into \SI{}{\nano\meter\per\hour} by scaling by the density of GaAs~\citep{weisburg,SATO1995508}.
Figure \ref{expectedsubrate} shows the expected sublimation rate into vacuum as a function of temperature.

The congruent temperature of GaP is \SI{571}{\degreeCelsius}~\citep{Amiri2022} therefore sublimation of GaP and GaAsP is also governed by the equilibrium pressure of gallium.
In this case, the sublimation rate of both materials should be similar.

The GaAsP region with index of refraction $n$ between the DBR interface and the photocathode surface forms a Fabry-Perot cavity~\citep{Fabry1899,fabry2}. 
This cavity stores light in the resonance condition
\begin{equation}
    2nL = m\lambda_R
    \label{resonancewave}
\end{equation}
where the product of $nL$ describes the optical path length, $m$ is an integer describing the order of the resonance, and $\lambda_R$ is the wavelength where a resonance occurs.
We used a numerical transfer matrix model (TMM)~\citep{Byrnes2016MultilayerOC} to simulate the reflectivity of the sample.
Here, \SI{1}{\nano\meter} of sublimated material provided \SI{0.74}{\nano\meter} of resonance shift.
While a \SI{0.74}{\nano\meter} shift is meaningful for fixed wavelength narrow bandwidth sources, the finesse of the cavity does not drastically change as material is sublimated, such that the enhancement in absorption the optical cavity provides can be recovered if it is possible to tune the source wavelength.

The samples were installed in a vacuum chamber behind an optical window to prevent oxidation during heating and to emulate the test environment used during operation. 
Additionally, the MOCVD sample was HCl etched using a \SI{20}{\percent} solution for \SI{30}{\second} prior to the mounting process. 
The MBE grown material contained an arsenic cap which we removed in vacuum after installation. In order to heat the cathode surface after installation to vacuum, we placed a heater rod on the back of the photocathode. 
Due to the separation between the heat source and the photocathode surface, we calibrated the surface temperature to within \SI{5}{\degreeCelsius}. 
A heat controller was used to maintain temperature over chosen durations.
We heated each sample in \SI{1.5}{\hour} increments for \SI{7.5}{\hour} total. 
We chose \SI{550}{\degreeCelsius} to observe sublimation because the expected resonance shift rate should exceed \SI{0.5}{\nano\meter\per\hour}. 
Before and after each heat, we measured the reflectivity spectra to extract the rate of resonance shift as a function of time heated.

We measured the reflectivity of both samples using an external reflectometer. 
Reflected power can be picked-off if the polarization of the light is rotated \SI{90}{\degree}. 
In the lab frame, we achieved this rotation by first transmitting light with polarization in the plane of incidence (POI) through the Glan-Laser polarizer before a zero-order quarter wave plate (QWP)~\citep{thorlabs:waveplate}. 
We oriented this QWP at +\SI{45}{\degree} with respect to the fast axis to produce right circularly polarized light before reaching vacuum window. 
When this light is reflected inside the chamber it produces left circularly polarized light that becomes polarized perpendicular to the POI after traveling back through the QWP. 
The polarization of the reflected light is rotated \SI{90}{\degree}, causing the Glan-Laser polarizer to reflect the light into a power meter. 
We recorded power before the vacuum window and after reflection to compute reflectivity.

We repeated this process over a range of wavelengths using a superK EXTREME supercontinuum white light source and a superK VARIA variable wavelength tunable filter with a \SI{2}{\nano\meter} bandwidth to locate the wavelength with minimum reflectivity~\citep{SuperKExtreme}.
Light from the reflectometer set-up was provided to the sample over a range of wavelengths from \SI{700}{\nano\meter} to \SI{800}{\nano\meter} in order to capture the resonance wavelength.

\begin{figure*}[ht]
\begin{minipage}{\columnwidth}
\begin{tikzpicture}[receive]
\begin{axis}
[
    legend style={font=\small},
    legend pos=north east,
    legend columns=6
    legend cell align=left,
    width=1\linewidth,
    height=.612\linewidth,
    xlabel={Wavelength (nm)},
    ylabel={Reflectivity (\%)},
    ymin=0, ymax=75,
    ytick={0, 15, 30, 45, 60, 75},
    xmin=780, xmax=810,
    tick label style={font=\footnotesize},
    label style={font=\footnotesize},
    name=border
]
\addplot [only marks, red, mark options={fill=red, draw=red}] table[x, y index=1] {data/MOCVD_raw1.dat};
\addplot [only marks, orange, mark options={fill=orange, draw=orange}] table [x expr=\thisrow{x}, y expr=\thisrow{y}-3] {data/MOCVD_raw2.dat};
\addplot [only marks, yellow!80!orange, mark options={fill=yellow!80!orange, draw=yellow!80!orange}] table [x expr=\thisrow{x}, y expr=\thisrow{y}+2] {data/MOCVD_raw3.dat};
\addplot [only marks, green!50!black, mark options={fill=green!50!black, draw=green!50!black}] table [x expr=\thisrow{x}, y expr=\thisrow{y}-13] {data/MOCVD_raw4.dat};
\addplot [only marks, blue, mark options={fill=blue, draw=blue}] table [x expr=\thisrow{x}, y expr=\thisrow{y}-17] {data/MOCVD_raw5.dat};
\addplot [only marks, red!60!blue, mark options={fill=red!60!blue, draw=red!60!blue}] table [x expr=\thisrow{x}, y expr=\thisrow{y}-21] {data/MOCVD_raw6.dat};
\legend{\SI{0}{\hour},\SI{1.5}{\hour},\SI{3}{\hour},\SI{4.5}{\hour},\SI{6}{\hour},\SI{7.5}{\hour}}

\end{axis}

\end{tikzpicture}

\end{minipage}
\hfill
\begin{minipage}{\columnwidth}
\begin{tikzpicture}[receive]
\begin{axis}
[
    legend style={font=\small},
    legend pos=north east,
    legend columns=6
    legend cell align=left,
    width=1\linewidth,
    height=.612\linewidth,
    xlabel={Wavelength (nm)},
    ylabel={Reflectivity (\%)},
    ymin=0, ymax=75,
    ytick={0, 15, 30, 45, 60, 75},
    xmin=750, xmax=770,
    tick label style={font=\footnotesize},
    label style={font=\footnotesize},
    name=border
]
\addplot [only marks, red, mark options={fill=red, draw=red}] table [x expr=\thisrow{x}, y expr=\thisrow{y}+10] {data/MBE_raw1.dat};
\addplot [only marks, orange, mark options={fill=orange, draw=orange}] table [x expr=\thisrow{x}, y expr=\thisrow{y}+4] {data/MBE_raw2.dat};
\addplot [only marks, yellow!80!orange, mark options={fill=yellow!80!orange, draw=yellow!80!orange}] table [x expr=\thisrow{x}, y expr=\thisrow{y}-2.5] {data/MBE_raw3.dat};
\addplot [only marks, green!50!black, mark options={fill=green!50!black, draw=green!50!black}] table [x expr=\thisrow{x}, y expr=\thisrow{y}-9.5] {data/MBE_raw4.dat};
\addplot [only marks, blue, mark options={fill=blue, draw=blue}] table [x expr=\thisrow{x}, y expr=\thisrow{y}-16] {data/MBE_raw5.dat};
\addplot [only marks, red!60!blue, mark options={fill=red!60!blue, draw=red!60!blue}] table [x expr=\thisrow{x}, y expr=\thisrow{y}-22] {data/MBE_raw6.dat};
\legend{\SI{0}{\hour},\SI{1.5}{\hour},\SI{3}{\hour},\SI{4.5}{\hour},\SI{6}{\hour},\SI{7.5}{\hour}}
\end{axis}

\end{tikzpicture}

\end{minipage}
\caption{Experimental reflectivity measurements for the MOCVD (LEFT) sample and MBE (RIGHT) sample around the primary resonance. Fabrication variability and wafer location alter the beginning resonance wavelength. Subsequent heat curves are offset to show the shift in minimum wavelength correlated to GaAs sublimation. The magnitude of reflectivity varied by less than \SI{5}{\percent} for both samples.}
\label{fig:sublimation}
\end{figure*}
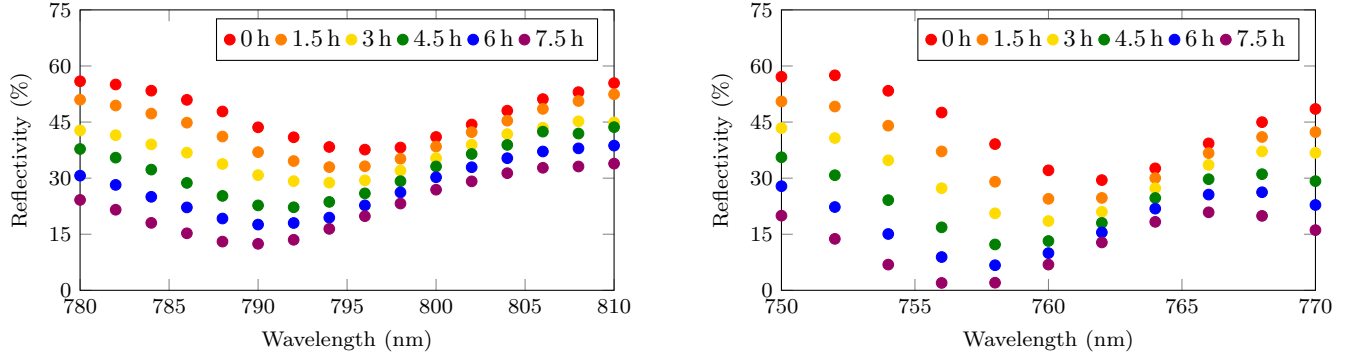

\begin{figure}[ht!]
\begin{tikzpicture}[receive]
\begin{groupplot}[
    group style={
        group size=1 by 2,
        vertical sep=10pt
    },
    width=1\columnwidth,
    height=.5\linewidth,
    xmin=0, xmax=8
]
\nextgroupplot[
    xticklabel=\empty,
    ylabel={Min Ref $\lambda$ (\SI{}{\nano\meter})},
    ymin=788, ymax=796,
]
\addplot[only marks, black] table[x=x, y=y, y error=err] {data/MOCVD_peakfit.dat};
\addplot[no marks, red, dashed, line width=2pt, mark options={fill=black, draw=black}] table[x, y index=1] {data/MOCVD_fitfunc1.dat};
\addplot[no marks, red, dashed, line width=2pt, mark options={fill=black, draw=black}] table[x, y index=1] {data/MOCVD_fitfunc2.dat};

\nextgroupplot[
    ylabel={Min Ref $\lambda$ (\SI{}{\nano\meter})},
    ymin=756, ymax=764,
    xlabel={Heat Duration (\SI{}{\hour})}
]
\addplot[only marks, black] table[x=x, y=y, y error=err] {data/MBE_peakfit.dat};
\addplot[no marks, red, dashed, line width=2pt, mark options={fill=black, draw=black}] table[x, y index=1] {data/MBE_fitfunc.dat};

\end{groupplot}
\end{tikzpicture}
\caption{The wavelength of minimum reflectivity as a function of time spent at \SI{550}{\degreeCelsius} for the MOCVD-grown sample (TOP) and MBE-grown sample (BOTTOM). A linear fit extracts the rate of change in the FP resonance location. For MOCVD, two linear fits are shown for before and after the position change.}
\label{fig:rate}
\end{figure}
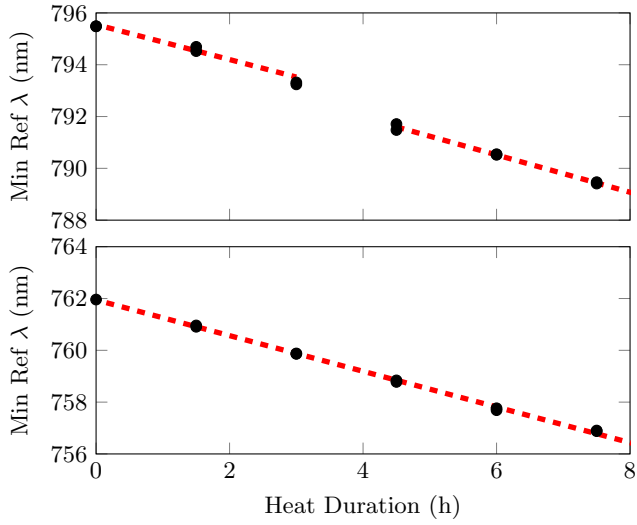

Fig. \ref{fig:sublimation} (LEFT) shows the reflectivity spectrum after each heat for the MOCVD sample and Fig. \ref{fig:sublimation} (RIGHT) for the MBE sample.
The magnitude of reflectivity at the minimum ranged from \SI{27}{\percent} to \SI{37}{\percent} for the MOCVD sample and \SI{19}{\percent} to \SI{24}{\percent} for the MBE sample.
We extracted the minimum of the reflectivity spectra using a local peak fit over a \SI{10}{\nano\meter} range.
We first recorded reflectivity at \SI{24}{\degreeCelsius} and again after roughly \SI{1}{\hour} at room temperature, typically 21-\SI{22}{\degreeCelsius}.
Since the cavity resonance depends on temperature through the index of refraction of the spacer material, and more weakly through its thermal expansion we applied an empirical correction of \SI{0.12}{\nano\meter} to \SI{0.36}{\nano\meter} to reference all resonance measurements to \SI{21}{\degreeCelsius}.
Figures \ref{fig:rate} (TOP) and \ref{fig:rate} (BOTTOM) plot the minimum reflectivity wavelengths with a linear fit for MOCVD and MBE respectively.

The FP resonance location in Fig. \ref{fig:rate} is extracted after heating for a duration of \SI{1.5}{\hour} at \SI{550}{\degreeCelsius}.
We evaluated the rate for the MOCVD sample at two locations that show consistent results.
The first fit provided a shift rate, $R_\lambda$ of \SI{0.68(03)}{\nano\meter\per\hour} and the second of \SI{0.72(03)}{\nano\meter\per\hour} which agree within uncertainty.
Table \ref{TOTALS} reports the average of these two values, \SI{0.70(02)}{\nano\meter\per\hour} for comparison with the other methods.
The MBE sample produced a similar shift rate, $R_\lambda$ of \SI{0.69(01)}{\nano\meter\per\hour}.

The TMM calculation allows $R_\lambda$ to be converted into $R_d$.
This conversion produces a measured sublimation rate, $R_d$, of \SI{0.94(03)}{\nano\meter\per\hour} for the MOCVD sample and \SI{0.93(01)}{\nano\meter\per\hour} for the MBE sample.

\begin{figure}[ht!]
    \begin{tikzpicture}[receive]
\begin{axis}
[
    legend style={font=\small},
    legend pos=north east, 
    width=1\linewidth,
    height=.612\linewidth,
    xlabel={Measurement during activation (\#)},
    ylabel={Min Ref $\lambda$ (\SI{}{\nano\meter})},
    ymin=780, ymax=782,
    xmin=0, xmax=10,
    tick label style={font=\footnotesize},
    label style={font=\footnotesize},
    error bars/error bar style={black},
    error bars/y dir=both,
    error bars/y explicit,
    name=border
]
\addplot[only marks, black] table[x=x, y=y, y error=err] {data/nochangeact.dat};

\end{axis}
\end{tikzpicture}
\caption{The wavelength of minimum reflectivity over a series of measurements during a full activation cycle. The wavelength remains fixed despite the ion flux hitting the surface.}
\label{fig:nochange}
\end{figure}
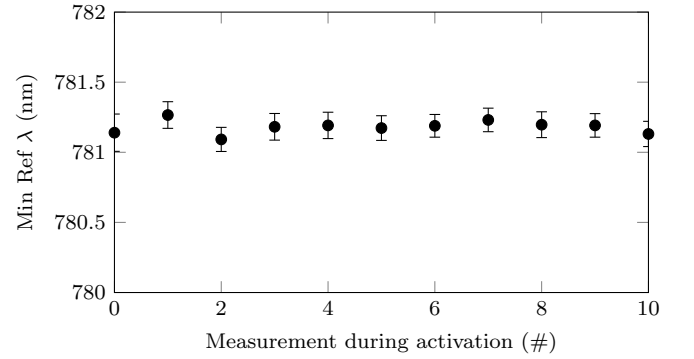

Additionally, during the activation process~\citep{GUO201765} for normal operations a photocathode can be exposed to a relatively high flux of ion-back-bombardment~\citep{yoskow}.
To confirm that sublimation is the primary cause of the FP resonance shifts, we measured the reflectivity multiple times during an activation.
Fig. \ref{fig:nochange} shows the minimum reflectivity wavelength across a 2 hour activation of an additional MOCVD sample where the location does not change.

\begin{figure*}
    \begin{tikzpicture}[font=\ttfamily]
    \node[inner sep=8pt] () at (0,0)
    {\includegraphics[width=0.965\textwidth]{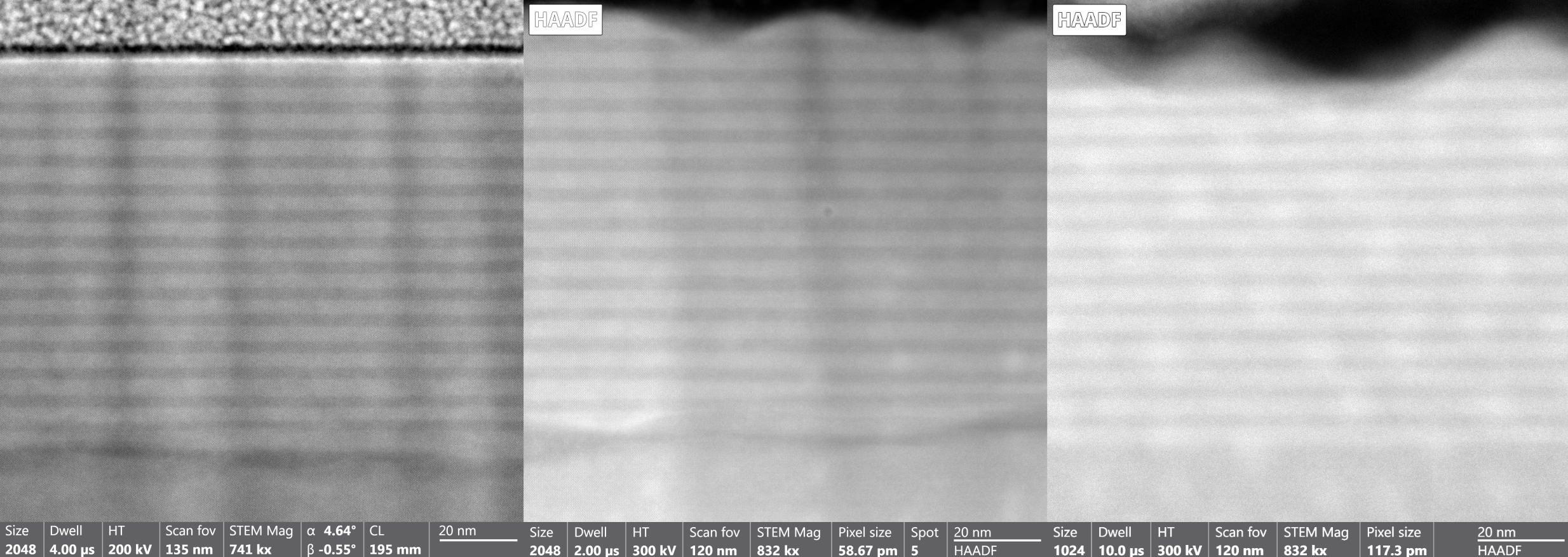}};
    \draw (0, -2.4) node[draw, fill=white, rectangle, rounded corners=0mm, minimum width=17cm, minimum height=0.5cm, inner sep=0] (sample) {};
    \draw (-6,-2.15) node[anchor=north, align=center, black] {\textbf{No exposure}};
    \draw (0,-2.15) node[anchor=north, align=center, black] {\textbf{MOCVD \SI{7.5}{\hour} at \SI{550}{\degreeCelsius}}};
    \draw (6,-2.15) node[anchor=north, align=center, black] {\textbf{MBE \SI{7.5}{\hour} at \SI{550}{\degreeCelsius}}};
    \fill[yellow] (-0.82,2.6) circle (2pt);
    \fill[yellow] (0.3,2.9) circle (2pt);
    \fill[yellow] (6.35,2.15) circle (2pt);
    \fill[yellow] (8.05,2.75) circle (2pt);
    
    \draw (8.3,-1.55) node[anchor=north, align=center, black] {\scriptsize \textbf{1}};
    \draw (8.3,-1.25) node[anchor=north, align=center, black] {\scriptsize \textbf{2}};
    \draw (8.3,-0.9) node[anchor=north, align=center, black] {\scriptsize \textbf{3}};
    \draw (8.3,-0.6) node[anchor=north, align=center, black] {\scriptsize \textbf{4}};
    \draw (8.3,-0.3) node[anchor=north, align=center, black] {\scriptsize \textbf{5}};
    \draw (8.3,-0.0) node[anchor=north, align=center, black] {\scriptsize \textbf{6}};
    \draw (8.3,0.325) node[anchor=north, align=center, black] {\scriptsize \textbf{7}};
    \draw (8.3,0.65) node[anchor=north, align=center, black] {\scriptsize \textbf{8}};
    \draw (8.3,0.95) node[anchor=north, align=center, black] {\scriptsize \textbf{9}};
    \draw (8.3,1.25) node[anchor=north, align=center, black] {\scriptsize \textbf{10}};
    \draw (8.3,1.6) node[anchor=north, align=center, black] {\scriptsize \textbf{11}};
    \draw (8.3,1.925) node[anchor=north, align=center, black] {\scriptsize \textbf{12}};
    \draw (8.3,2.225) node[anchor=north, align=center, black] {\scriptsize \textbf{13}};
    \draw (8.3,2.575) node[anchor=north, align=center, black] {\scriptsize \textbf{14}};

    \draw (2.5,-1.275) node[anchor=north, align=center, black] {\scriptsize \textbf{1}};
    \draw (2.5,-1.05) node[anchor=north, align=center, black] {\scriptsize \textbf{2}};
    \draw (2.5,-0.75) node[anchor=north, align=center, black] {\scriptsize \textbf{3}};
    \draw (2.5,-0.4) node[anchor=north, align=center, black] {\scriptsize \textbf{4}};
    \draw (2.5,-0.05) node[anchor=north, align=center, black] {\scriptsize \textbf{5}};
    \draw (2.5,0.25) node[anchor=north, align=center, black] {\scriptsize \textbf{6}};
    \draw (2.5,0.6) node[anchor=north, align=center, black] {\scriptsize \textbf{7}};
    \draw (2.5,0.925) node[anchor=north, align=center, black] {\scriptsize \textbf{8}};
    \draw (2.5,1.25) node[anchor=north, align=center, black] {\scriptsize \textbf{9}};
    \draw (2.5,1.55) node[anchor=north, align=center, black] {\scriptsize \textbf{10}};
    \draw (2.5,1.875) node[anchor=north, align=center, black] {\scriptsize \textbf{11}};
    \draw (2.5,2.225) node[anchor=north, align=center, black] {\scriptsize \textbf{12}};
    \draw (2.5,2.575) node[anchor=north, align=center, black] {\scriptsize \textbf{13}};
    \draw (2.5,2.875) node[anchor=north, align=center, black] {\scriptsize \textbf{14}};
    \end{tikzpicture}
\caption{STEM images comparing an undamaged MOCVD sample to the two samples heated in this study (scale bar is 20 nm for all images). For both the MOCVD and MBE sample significant damage is observed from sublimation. The magnitude of sublimation was found by counting the number of pairs from the beginning of the SSL and assigning a known pair thickness. The minimum and maximum sublimation location is shown in yellow.}
\label{temimage}
\end{figure*}

Finally, in order to confirm the hypothesis that material sublimated, we removed and analyzed the samples by STEM.
Fig. \ref{temimage} compares these results to an unheated MOCVD sample; both heated samples exhibit sublimation.

Table \ref{TOTALS} compares the sublimation rates measured by the two different techniques.
We obtained the STEM sublimation rates $R_d$ using digital image analysis to extract the minimum and maximum amount of sublimated material by counting the number of pairs and assigning the design pair thickness.
The minimum and maximum are averaged to assign an amount of material sublimated and the difference from the average assigns the error.
We divided this average thickness of material removed by the \SI{7.5}{\hour} heating duration of each sample to produce the rate shown in Table \ref{TOTALS}.

\begin{table}[h!]
\centering
\begin{tabular}{c|c|c|c}
Method  &  Device    & $R_\lambda$ (\SI{}{\nano\meter\per\hour}) & $R_d$ (\SI{}{\nano\meter\per\hour}) \\ \hline
OPTICAL SHIFT& MOCVD & 0.70 $\pm$ 0.02  & 0.94 $\pm$ 0.03          \\ \hline
OPTICAL SHIFT & MBE  & 0.69 $\pm$ 0.01  & 0.93 $\pm$ 0.01          \\ \hline
STEM & MOCVD    & n/a                & 1.0 $\pm$ 0.2          \\ \hline
STEM & MBE      & n/a                & 1.0 $\pm$ 0.4          
\end{tabular}
\caption{The FP resonance shift rate, $R_\lambda$, and the sublimation rate, $R_d$, for each measurement method.}
\label{TOTALS}
\end{table}

The surface temperature is inferred from a thermocouple measurement a distance away from the sample, so an offset of \SI{1}{\percent} could be feasible for the system.
At \SI{550}{\degreeCelsius} the calculated Langmuir sublimation rate, $R_d$, is \SI{0.73(19)}{\nano\meter\per\hour}, which is in agreement with the experimental results.
This slightly lower rate suggests the actual surface temperature was slightly greater than \SI{550}{\degreeCelsius}.

This shift in the FP resonance has many applications beyond the measurement of the surface damage during preparation.
For example, the FP resonance location may be a useful metric to observe operational damage from high-energy ions in photoinjectors.

During regular operation of a spin-polarized physics program GaAs based photocathodes are often heated at roughly \SI{350}{\degreeCelsius} and \SI{520}{\degreeCelsius}.
At \SI{350}{\degreeCelsius} the rate of sublimation is negligible as seen in Fig. \ref{expectedsubrate}; however, \SI{520}{\degreeCelsius} begins to sublimate at around \SI{0.2}{\nano\meter\per\hour}.
This sublimation poses a challenge for DBR photocathodes because after sublimation the desired enhancement in optical absorption corresponding to the FP resonance will move away from the expected wavelength of the light source.
A reduction in the maximum photocurrent following every heat treatment is not ideal for prolonged operation required in most accelerator programs.
Additionally, many photocathodes contain a highly doped surface layer intended to reduce surface charge limit (SCL)~\citep{wanglifetime}.
If sublimation removes this region, the mitigation of SCL will be reduced which could cause decreased performance.
As we observed sublimation to also roughen the surface, the emittance of the beam could grow following extended heat cycles~\citep{emittance1,emittance2}.

We correlated shifts in the FP resonances with the sublimation rate of both MOCVD-grown and MBE-grown SSL GaAs based photocathodes with sub-nm sensitivity at a temperature of roughly \SI{550}{\degreeCelsius}.
A simple reflectometer measured the wavelength dependent reflectivity spectra during more than \SI{7}{\hour} of heat exposure.
We normalized the reflectivity minimum, corresponding to enhanced absorption, to the sublimation rate using a TMM calculation for variable material loss.
The MOCVD sample yielded a sublimation rate of (0.94 $\pm$ 0.03) \SI{}{\nano\meter\per\hour} while the MBE sample yielded a rate of (0.93 $\pm$ 0.01) \SI{}{\nano\meter\per\hour}.
We isolated resonance shifts to sublimation from heat treatment by observing no shift across multiple measurements during a photocathode activation.
STEM measurements support the experimentally determined sublimation rate from the FP resonances.
Overall, this technique provides a non-destructive measurement of surface damage in situ which can be applied directly to the operational environment regardless of the materials being used or the structure of the devices.

\begin{acknowledgments}
This material is based upon work supported by the U.S. Department of Energy, Office of Nuclear Physics, Contract Number DE-SC0025519.
This material is based upon work supported by the U.S. Department of Energy, Office of Science, Office of Nuclear Physics under Contract No. 89243126CSC000213.
This work was performed in part at the Analytical Instrumentation Facility (AIF) at North Carolina State University, which is supported by the State of North Carolina and the National Science Foundation (award number ECCS-2025064). 
The AIF is a member of the North Carolina Research Triangle Nanotechnology Network (RTNN), a site in the National Nanotechnology Coordinated Infrastructure (NNCI).
\end{acknowledgments}

\section*{Author Declarations}
\noindent \textbf{Conflict of Interest}

The authors have no conflicts to disclose.

\section*{Data Availability Statement}

The data that support the findings of this study are available from the corresponding author upon reasonable request.

\bibliography{apssamp}

\end{document}